\documentclass[reprint,superscriptaddress,nofootinbib,longbibliography]{revtex4-2}
\usepackage{graphicx} 
\usepackage{microtype}
\usepackage[english]{babel}
\usepackage[utf8]{inputenx}
\usepackage{wrapfig}
\usepackage{float}
\usepackage{ifthen,pifont,epsfig,latexsym,amsmath,amssymb}
\usepackage[strict]{changepage}
\usepackage{graphicx}
\usepackage{amsmath}
\usepackage{amsfonts}
\usepackage[T1]{fontenc}
\usepackage{listings}
\usepackage{braket}
\usepackage{xcolor}
\usepackage[margin=2cm]{geometry}
\usepackage{comment}
\usepackage{physics}
\usepackage{breqn}
\usepackage{bm}
\usepackage{url}
\usepackage{orcidlink}

\usepackage{etoolbox}

\makeatletter
\patchcmd{\doauthor}
  {\unskip\@listcomma}
  {\unskip}
  {}
  {\PackageWarningNoLine{author-list}{Could not remove the original comma}}
\patchcmd{\doauthor}
  {\space\@listand}
  {\@listcomma\space\@listand}
  {}
  {\PackageWarningNoLine{author-list}{Could not move the comma after the superscripts}}
\makeatother

\usepackage{hyperref}
 \hypersetup{
 colorlinks = true, 
 urlcolor = blue, 
 linkcolor = blue, 
 citecolor = red 
 }
\newcommand{\bfk}{\mathbf{k}}

\newcommand{\INFN}{INFN - Sezione collegata di Salerno, Complesso Univ. Monte S. Angelo, I-80126 Napoli, Italy}

\newcommand{\UNISA}{Physics Department ``E.R. Caianiello'', Universit\`a degli Studi di Salerno, Via Giovanni Paolo II, 132, I-84084 Fisciano (Sa), Italy}

\newcommand{\CNR}{CNR-SPIN, I-84084 Fisciano (Salerno), Italy, c/o Universit\`a di Salerno, I-84084 Fisciano (Salerno), Italy}

\newcommand{\HALLE}{ Institute of Physics, Martin-Luther-Universität Halle-Wittenberg and Halle-Berlin-Regensburg Cluster of Excellence ``Center for Chiral Electronics'', 06120 Halle (Saale), Germany}

\newcommand{\MAXPLANK}{Max Planck Institute of Microstructure Physics, Weinberg 2, 06120 Halle (Saale), Germany}

\begin{document}

\title{Multiorbital driving of quantum geometric nonlinear transport in oxide interfaces
}

\author{Irene Gaiardoni\,\orcidlink{0009-0007-8348-3405}}
\email{igaiardoni@unisa.it}
\affiliation{\UNISA}

\author{Mattia Trama\,\orcidlink{0000-0001-5032-7974}}
\email{mtrama@unisa.it}
\affiliation{\UNISA}
\affiliation{\INFN}

\author{Roberta Citro\,\orcidlink{0000-0002-3896-4759}}
\email{rocitro@unisa.it}
\affiliation{\UNISA}
\affiliation{\INFN}
\affiliation{\CNR}

\author{Annika Johansson\,\orcidlink{0000-0001-5453-6693}}
\email{annika.johansson@physik.uni-halle.de}
\affiliation{\HALLE}
\affiliation{\MAXPLANK}

\begin{abstract}
Quantum geometry plays a relevant role in determining the intrinsic non-linear transport properties of low-dimensional electron systems. Here, we theoretically investigate the quantum geometric contribution to the non-linear conductivity in $\text{SrTiO}_3$-based two-dimensional electron gases, comparing the prototypical (001) and (111) crystallographic interface orientations. 
Using a multi-orbital tight-binding approach, we uncover the crucial role of the multiband electronic structure in determining the quantum-geometric contribution to nonlinear transport.
In particular, we show how the hybridization of $t_{2g}$ orbitals near avoided band crossings shapes the momentum-space distribution of the band-normalized quantum metric and the resulting nonlinear conductivity.
Furthermore, we reveal a strong dependence on crystal orientation: while the (001) interface exhibits sharp peak-and-dip features confined around Lifshitz transitions, the trigonal symmetry of the (111) interface promotes strong orbital mixing. 
Our results highlight how the interplay between orbital hybridization and crystallographic orientation gives rise to distinct quantum-geometric nonlinear transport responses in oxide interfaces.
\end{abstract}

\maketitle

\section{Introduction}
Nonlinear electronic transport has emerged as a powerful probe of the electronic properties of quantum materials, including magnetic topological systems and spin-orbit-coupled oxide interfaces~\cite{wang2023quantum,sala2025the, ma2019observation, kang2019nonlinear, huang2023giant, tiwari2021giant, ma2022growth, shvetsov2019nonlinear, kumar2021room, dzsaber2021giant}. In particular, second-order transport responses can reveal intrinsic contributions governed by the momentum-space geometry of Bloch wave functions, providing information beyond that encoded in the electronic band dispersion alone~\cite{kaplan2024unification,jiang2025revealing, resta2011insulating, torma2023essay, wang2021intrinsic, liu2021intrinsic, gao2023quantum}.\\
Such geometric information is encoded in the quantum geometric tensor 
\begin{equation}
 \mathcal{Q}^n_{ij}= \sum_{m \neq n} \mathcal{A}_{nm}^a \mathcal{A}_{mn}^b
 \label{eq1}
 \end{equation}
 with $\mathcal{A}_{nm}^a= \bra{n}i \partial_a \ket{m}$  being the interband Berry connection, associated with the cell-periodic Bloch states $\ket{u_{n\mathbf{k}}}$~\cite{provost1980riemannian,kolodrubetz2017geometry}. Its imaginary part is related to the well-known Berry curvature ($\Omega^n_{ij}=-2{\rm Im}\,\mathcal{Q}^n_{ij}$), whose flux through a surface determines the geometric phase acquired along its boundary in momentum space and underlies a wide range of linear and nonlinear transport phenomena~\cite{xiao2010berry,sodemann2015quantum, du2018band, du2019disorder}.\\
The real part, $g^n_{ij}={\rm Re}\,\mathcal{Q}^n_{ij}$, defines the quantum metric, which provides a gauge-invariant measure of how rapidly the Bloch states vary with crystal momentum. More precisely, it quantifies the infinitesimal distance between neighboring quantum states through the reduction of their squared overlap~\cite{provost1980riemannian,kolodrubetz2017geometry,yu2025quantum},

\[
|\langle u_{n\mathbf{k}}|u_{n,\mathbf{k}+d\mathbf{k}}\rangle|^2
=
1-g_{ij}^{n}(\mathbf{k})\,dk_i\,dk_j+\mathcal{O}(dk^3).
\]
While this distance is defined in terms of the overlap between states belonging to the same band at different momenta, its value inherently reflects the multiband structure of the system. Indeed, as evident from Eq.~\eqref{eq1}, the momentum dependence of a given Bloch state is governed by its interband connections with the other eigenstates
and can therefore be enhanced when the orbital character of the Bloch states changes rapidly,
as occurs near avoided crossings.
Its integral over the Brillouin zone is also related to the spatial spread of the corresponding Wannier functions~\cite{verma2026quantum}. 
\begin{table*}[t]
    \caption{
    Symmetry constraints on the in-plane nonlinear conductivity tensor,
    defined by $j_i^{(2)}=\sigma_{ijk}E_jE_k$, with
    $\sigma_{ijk}=\sigma_{ikj}$. The fourth column lists the components
    allowed by the unitary spatial symmetries. The last column refers
    specifically to the time-reversal-odd contribution generated by the
    dipole of the BNQM. For the (111) orientation,
    $x\parallel[\bar{1}10]$ and $y\parallel[\bar{1}\bar{1}2]$.
    }
    \label{tab:symm}
    \begin{ruledtabular}
        \begin{tabular}{ccccc}
            Orientation
            & Zeeman field
            & Residual symmetry
            & Symmetry allowed components
            & BNQM-dipole components
            \\
            \hline
            $(001)$
            & $B=0$
            & $C_{4v}+\mathcal T$
            & None
            & None
            \\
            $(111)$
            & $B=0$
            & $C_{3v}+\mathcal T$
            &
            $\sigma_{xxy}=\sigma_{xyx}
            =\sigma_{yxx}=-\sigma_{yyy}$
            & None
            \\
            $(001)$
            & $B_y\neq0$
            &
            $\{1,M_{xz},\mathcal TC_{2z},\mathcal TM_{yz}\}$
            &
            $\sigma_{xxx},\ \sigma_{xyy},\
            \sigma_{yxy}=\sigma_{yyx}$
            &
            $\sigma_{xxx}^{\rm QG},\ \sigma_{xyy}^{\rm QG},\
            \sigma_{yxy}^{\rm QG}=\sigma_{yyx}^{\rm QG}$
            \\
            $(111)$
            & $B_y\neq0$
            & $\{1,\mathcal TM_{yz}\}$
            & All six components
            &
            $\sigma_{xxx}^{\rm QG},\ \sigma_{xyy}^{\rm QG},\
            \sigma_{yxy}^{\rm QG}=\sigma_{yyx}^{\rm QG}$
        \end{tabular}
    \end{ruledtabular}
\end{table*}

Despite its mathematical nature, the quantum metric enters directly into nonlinear transport through the response of Bloch electrons to an applied electric field. Indeed the electric field induces virtual interband mixing and a second-order correction to the $n$-th band energy
as $\delta\varepsilon_{n\mathbf{k}}^{(2)}
=
e^2E_iE_jG_{ij}^n(\mathbf{k})$ with $E_i$ the $i$-th component of the electric field and
\begin{equation}
    G_{ij}^n(\mathbf{k})
    =
    2\hbar^2
    \sum_{m\neq n}
    \frac{
    {\rm Re}\left[
    \mel{u_{n\mathbf{k}}}{\hat v_i}{u_{m\mathbf{k}}}
    \mel{u_{m\mathbf{k}}}{\hat v_j}{u_{n\mathbf{k}}}
    \right]
    }
    {\left(\varepsilon_{n\mathbf{k}}
    -\varepsilon_{m\mathbf{k}}\right)^3},
    \label{eq:BNQM}
\end{equation}
being the band normalised quantum metric (BNQM), $\hat{\mathbf{v}}= \frac{1}{\hbar}\frac{\partial}{\partial \mathbf k} \hat{H}$ the velocity operator, and $\hat{H}$ the Hamiltonian. 
Therefore, the derivative of this correction modifies the wave-packet velocity as $\delta {v^n_{a}}^{(2)}\propto E_iE_j\partial_{k_a}G^n_{ij}$, giving rise to an intrinsic nonlinear current governed by the dipole of the BNQM~\cite{kaplan2024unification,das2023intrinsic}.
\begin{figure}
    \centering
    \includegraphics[width=0.99\linewidth]{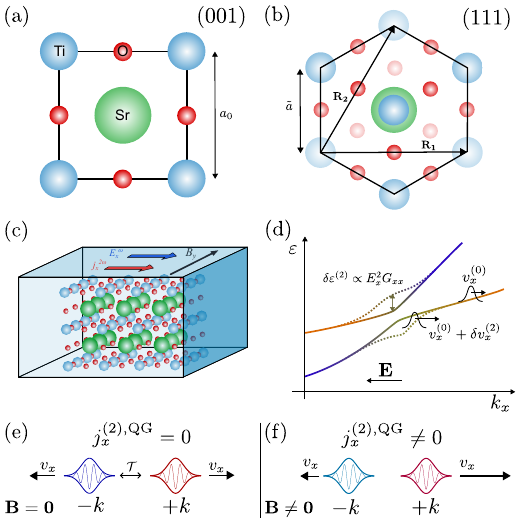}
    \caption{Schematic top-view of the (a) (001) and (b) (111) STO lattice geometry. $a_0=3.905\text{\AA}$ is the lattice constant, while $\tilde{a}=\sqrt{2/3} \ a_0$ is the in-plane lattice parameter of the triangular lattice; $\mathbf{R}_1=\tilde{a}\sqrt{3}\mathbf{\hat{x}}$ and $\mathbf{R}_2=\tilde{a}\sqrt{3}/2(\mathbf{\hat{x}}+\sqrt{3}\mathbf{\hat{y}})$ are the lattice vectors where $\mathbf{\hat{x}}\equiv[\bar{1}10]$ and $\mathbf{\hat{y}}\equiv[\bar{1}\bar{1}2]$ while considering the (111) interface. Blue, red, and green spheres represent Ti, O, and Sr atoms, respectively. (c) Schematic illustration of the nonlinear transport geometry: while an external magnetic field $B_y$ is applied, the system generates a second-harmonic intrinsic current in response to an electric field. (d-f) Schematic mechanism of the origin and effect of the quantum geometric contribution. (d) A second order energetic correction modifies the bands close to the already present avoided crossings modifying the group velocity. Because of time-reversal symmetry, in absence of magnetic field (e), the correction to the group velocity of two wave-packets at opposite $\mathbf{k}$ are the same; while in presence of $\bf B$ (f), the correction do not compensate and the intrinsic current $\mathbf{j}^{(2),{\rm QG}}$ does not vanish. }
    \label{fig:fig_0_sketch}
\end{figure}

Recently, oxide interfaces have emerged as suitable platform to investigate the quantum geometry in systems with strong spin-orbit coupling (SOC) and broken inversion symmetry, hosting Rashba SOC and the associated spin-momentum locking~\cite{sala2025the, caviglia2010tunable}. 
The key role of Rashba SOC in electronic transport has been widely investigated, revealing its influence on a variety of phenomena in normal and superconducting state ~\cite{caviglia2008electric, trama2022tunable,gaiardoni2025edelstein,SciPostPhys.17.4.101,trama2026inhomogeneous}, ranging from spin-charge interconversion to nonlinear and other spin-orbit-driven transport responses, both at oxide interfaces and in related systems ~\cite{zhai2023large,trama2025nonlinear,maiellaro2025theory,gaiardoni2026boltzmann,trama2023effect,trama2021straininduced,trama2022gate,SciPostPhys.20.5.144,varotto2022direct,leiva2023spin,johansson2016theoretical,2th8-nyz8,donofrio2026spinselectivityinducednoncollinear}. These studies further motivate the investigation of the role of Rashba SOC in quantum-geometric nonlinear transport, where its interplay with the multiband electronic structure can give rise to nontrivial transport responses.

An intrinsic contribution to the nonlinear in-plane magnetotransport was identified at the (111) LaAlO$_3$/SrTiO$_3$ (LAO/STO) interface upon breaking time-reversal symmetry with an in-plane magnetic field~\cite{sala2025the}. This response was attributed to the quantum metric associated with the Rashba SOC: 
while for zero magnetic field, the quantum metric-induced contribution to nonlinear charge  transport vanishes for symmetry reasons, a finite contribution occurs in the presence of an in-plane magnetic field (Tab.~\ref{tab:symm}).
While this symmetry-controlled mechanism can be captured within a minimal Rashba model~\cite{sala2025the}, a more realistic description of oxide interfaces requires accounting for their multiorbital electronic structure. Indeed, in the two-band Rashba model, the BNQM has equal magnitude and opposite sign in the two bands, both with and without an external magnetic field. While time-reversal symmetry enforces a vanishing nonlinear conductivity at zero field, breaking this symmetry allows a finite response by preventing the cancellation of momentum-space contributions. However, in multiorbital systems, the interplay between states with different orbital character and their hybridization gives rise to a richer electronic band structure, characterized by avoided band crossings. In the presence of an external in plane magnetic field, these features can play an important role in shaping the quantum-geometric contribution to nonlinear transport, motivating an investigation beyond the minimal Rashba model.\\
In this work, we investigate the role of these multiorbital effects in the quantum-geometric contribution to nonlinear magnetotransport at the (001) and (111) LAO/STO interfaces (Fig.~\ref{fig:fig_0_sketch}). Using multiorbital tight-binding models, we examine how the different electronic structures and orbital hybridization mechanisms influence the nonlinear response. We find that the (001) interface exhibits pronounced peak-and-dip structures associated with interorbital hybridization near Lifshitz transitions, whereas the (111) interface displays a sizable and nonmonotonic response governed by the interplay between its multiorbital band structure and orbital mixing.\\

\section{Model}
We describe the two STO-based interfaces through orientation-dependent
tight-binding Hamiltonians reproducing their low-energy conduction
bands. In both cases, the Hamiltonian can be schematically decomposed as
\begin{equation}
    H^{(\eta)}
    =
    H_{\rm KIN}^{(\eta)}
    +H_{\rm SOC}^{(\eta)}
    +H_{\rm dist}^{(\eta)}
    +H_{v}^{(\eta)} + H_{B},
\end{equation}
with $\eta=(001),(111)$ and the different terms account for the kinetic energy, atomic spin-orbit coupling, symmetry-lowering distortions, orbital mixing, and a Zeeman coupling, respectively.

The Hamiltonians of the two orientations differ in their low-energy orbital-layer basis.
Excluding spin, the (001) interface is described by a reduced four-band model containing the three $t_{2g}$ orbitals of the interfacial Ti layer and the lowest $d_{xy}$-derived subband associated with the neighboring layer.

For the (111) interface, we retain the three Ti $t_{2g}$ orbitals on each plane of a Ti bilayer, resulting in six orbital-layer degrees of freedom, of which only three Kramers doublets are occupied in the filling range considered here. 
In the absence of SOC and crystal-field splitting, the three orbital bands are degenerate at $\mathbf{k}=0$. The trigonal crystal field lifts this degeneracy into a lower $a_g$ singlet and an $e_g^\pi$ doublet.

\begin{figure*}[t]
\centering
\includegraphics[
width=0.98\textwidth]{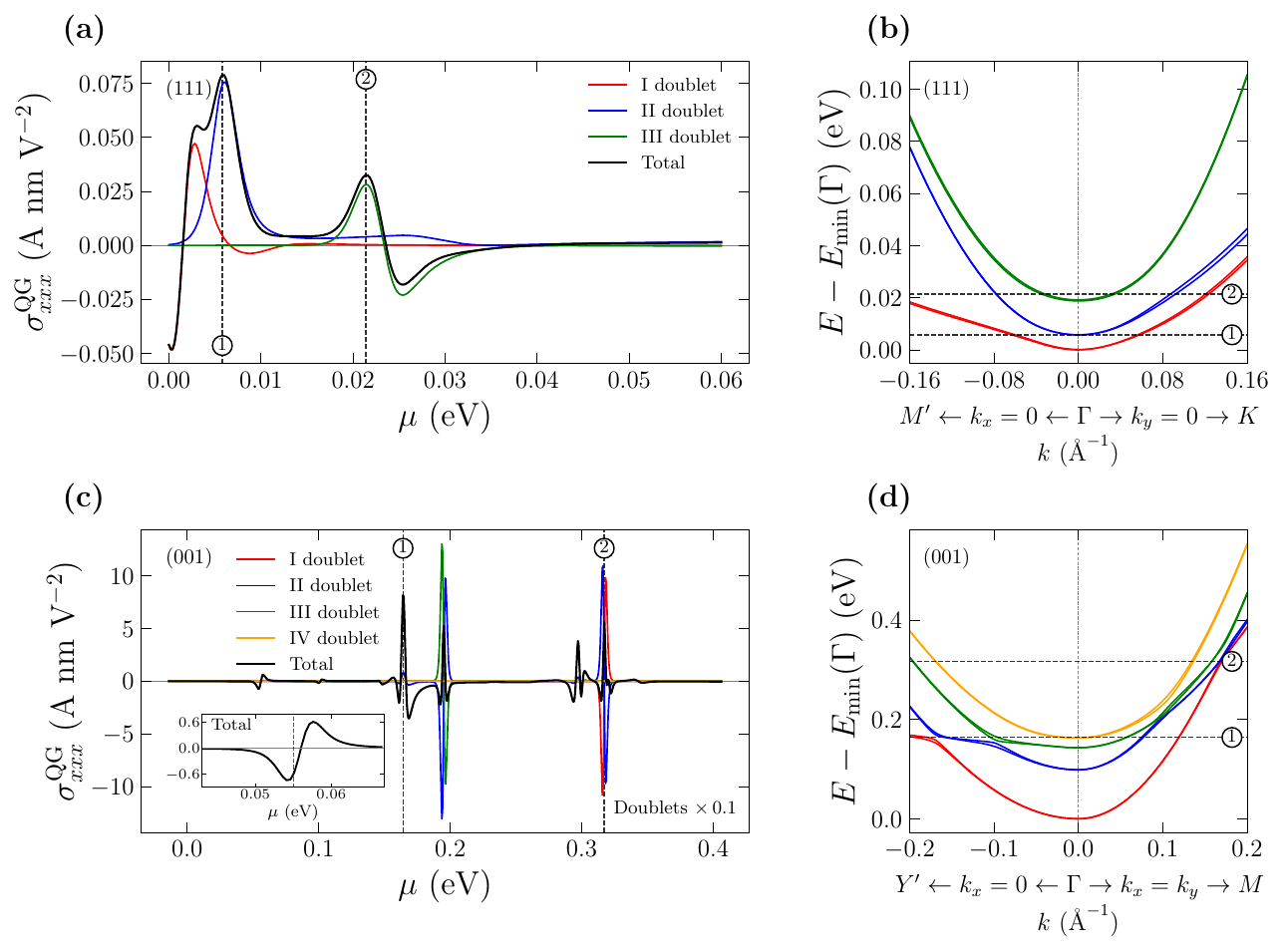}
\caption{
(a) Quantum-geometry-induced nonlinear conductivity,
$\sigma_{xxx}^{\mathrm{QG}}$, as a function of the chemical
potential for the (111) LAO/STO interface, resolved into the contributions of the first three doublets.
(b) Corresponding electronic band structure.
(c) Quantum-geometry-induced nonlinear conductivity,
$\sigma_{xxx}^{\mathrm{QG}}$, as a function of the chemical
potential for the (001) LAO/STO interface, together with the contributions of the individual doublets; the latter are rescaled by a factor of $0.1$ for clarity
(d) Corresponding electronic band structure.
The chemical potential and band energies are measured from the minimum of the lowest band at $\Gamma$. 
The labels 1 and 2 mark the chemical potentials selected for the momentum-resolved analysis. 
All calculations are performed for an in-plane magnetic field $B_y=5\,\mathrm{T}$ and $T=10\,\mathrm{K}$.
} 
\label{fig:band_structure}
\end{figure*}

The explicit Hamiltonian matrices and their parametrization are reported in Appendix~\ref{app:hamiltonians}. The resulting low-energy band structures at $B_y=5\,{\rm T}$ are shown in Figs.~\ref{fig:band_structure}(b) and \ref{fig:band_structure}(d).

Without Zeeman coupling, $H^{(001)}$ has $C_{4v}$ symmetry, whereas $H^{(111)}$ has $C_{3v}$ symmetry. An in-plane second-order response of the form $j_i^{(2)}=\sigma_{ijk}E_jE_k$ vanishes identically for $C_{4v}$.
By contrast, $C_{3v}$ allows a response characterized by a single independent conductivity coefficient $\sigma$, $j_x^{(2)}=2\sigma E_xE_y$, $j_y^{(2)}=\sigma(E_x^2-E_y^2)$.
Here, $x$ and $y$ always denote in-plane directions; for the (111) orientation, $x\parallel[\bar{1}10]$ and $y\parallel[\bar{1}\bar{1}2]$ (Tab.~\ref{tab:symm}). However, the second-order response allowed by $C_{3v}$ cannot originate from the dipole of the BNQM.
Indeed, time-reversal symmetry imposes
$G_{ij}^{n}(\mathbf{k})=G_{ij}^{n}(-\mathbf{k})$, making its momentum-space dipole odd in $\mathbf{k}$ and hence zero upon integration over the Brillouin zone. A Zeeman field along $y$ breaks time-reversal symmetry and further reduces the magnetic point group, thereby allowing a finite intrinsic BNQM-dipole response. The resulting constraints on the nonlinear conductivity tensor are summarized in Table~\ref{tab:symm}. 
In the following, we focus on the longitudinal configuration with the Zeeman field along $y$ and both the electric field and the induced current along $x$. We therefore evaluate $\sigma_{xxx}^{\rm QG}$, which is symmetry-allowed for both the (001) and (111) orientations.

In the Boltzmann semiclassical framework, the longitudinal nonlinear conductance reads

\begin{equation}
    \sigma^{QG}_{xxx}(\mu)=\frac{3e^3}{2\hbar}\sum_n \int \frac{d^2\bfk}{4\pi^2} \Lambda^n_{xxx}(\bfk) f_{T}(\varepsilon_n(\bfk)-\mu),
    \label{eq:nonlinearQG}
\end{equation}
$\Lambda_{xxx}^n(\bfk)=\partial_{k_x}G_{xx}^n(\bfk)$
where $f_{T}(x)=1/(1+e^{x/k_BT})$ is the Fermi-Dirac distribution function ~\cite{kaplan2024unification}. We notice that within this approximation Eq.~\eqref{eq:nonlinearQG} does not depend on the scattering time due to its intrinsic nature.

\section{Results}
Figs.~\ref{fig:band_structure}(a) and ~\ref{fig:band_structure}(c) show the quantum-geometric contribution to the nonlinear
conductivity as a function of the chemical potential $\mu$ for the (111) and (001)
LAO/STO interfaces at an in-plane magnetic field of $B=5$ T (the behavior of the electronic occupation on the filling is shown in Appendix~\ref{app:density}.). Although the
lowest-energy bands of both systems display an effective Rashba-like
splitting, their nonlinear responses are qualitatively different. 
\\For the (111) interface, panel (a) immediately reveals a strongly nonmonotonic dependence of the nonlinear charge conductivity on the chemical potential. As discussed above, the BNQM is controlled not only by the structure of the avoided crossings, but also by the matrix elements $\mel{u_{n\mathbf{k}}}{\hat v_i}{u_{m\mathbf{k}}}
    \mel{u_{m\mathbf{k}}}{\hat v_j}{u_{n\mathbf{k}}}
    $ entering its definition. Both ingredients must therefore be considered to fully understand the nonlinear response. In the absence of the trigonal crystal field, SOC splits the $t_{2g}$ manifold at $\Gamma$ into a $J=3/2$ quadruplet and a $J=1/2$ doublet~\cite{trama2022gate}. The first and third doublets have a similar $J_{111}=\pm1/2$ character, whereas the second one is mainly associated with the $J_{111}=\pm3/2$ sector. One might therefore expect the first and third doublets to display a similar behavior based on their $J_{111}$ component. However, panel (a) reveals distinct filling-dependent contributions from the individual doublets. At the lowest chemical potentials, the nonlinear response is mainly determined by the first doublet. As the chemical potential increases, the second doublet develops a pronounced contribution, while the third doublet becomes increasingly relevant and dominates the response at higher fillings.  Notably, the corresponding 
    response around the chemical potential marked as 2 in Fig.~\ref{fig:band_structure}(a) can be
    also captured by the effective two-band model of the (111) interface discussed in the literature \cite{sala2025the}.
    As discussed in Appendix~\ref{app:effective_model}, the avoided crossings at finite momentum give rise to a local quantum-geometric enhancement analogous to that of a simple Rashba model, although originating from interorbital hybridization.
    Interestingly, despite their similar spin-orbital character, the first and third doublets exhibit markedly different contributions to the nonlinear conductivity. This behavior highlights that the spin-orbital character of the states alone is not sufficient to determine the quantum-geometric response. The specific hybridization channels, the avoided-crossing structure, and the corresponding matrix-element weights are equally important in setting both the magnitude and the chemical-potential dependence of the response.\\
Regarding the (001) interface, the quantum-geometric response in
Fig.~\ref{fig:band_structure}(c) is strongly concentrated within narrow
ranges of chemical potential, in marked contrast with the broader filling
dependence found for the (111) orientation. This behavior reflects the
stronger role of confinement, which allows the different doublets to retain a predominantly well-defined orbital character over extended regions of momentum space. In particular, the orbital-mixing term of the Hamiltonian vanishes at $\Gamma$ and becomes finite away from the zone center. Moreover, its structure favors specific hybridization channels, directly coupling the $d_{xy}$-derived subbands to the $d_{yz}/d_{zx}$ sector, while no direct orbital-mixing term connects the two $d_{xy}$ subbands.
The two main regions where the total conductivity is strongly enhanced are associated with pronounced avoided crossings in the band structure
(Fig.~\ref{fig:band_structure}(d)). At lower energies,  for chemical potentials between $\mu= 0.15$ eV and $\mu=0.2$ eV, the dominant
hybridization mainly involves the second and third doublets, whereas the
higher-energy region predominantly involves the first and second ones.
Since a large BNQM requires both a small interband separation and sizeable interband matrix elements, these orbital hybridizations provide favorable conditions for a strong quantum-geometric response. In particular, the lower-energy region can be related to the hybridization between the predominantly $d_{xy2}$-like second doublet and the $d_{yz}/d_{zx}$-like third one, as we can see around $\mu=0.2$ eV. 
Across this avoided crossing, the second doublet acquires a sizeable $d_{yz}/d_{zx}$ component, which can in turn enhance its hybridization with the predominantly $d_{xy1}$-like first doublet in the higher-energy region.
The pronounced peak marked by $1$ is also  enhanced  by the small energy separation between bands of the second doublets (see blue curve in Fig.~\ref{fig:band_structure}~(c) and (d)), suggesting that intra-doublet energy splitting plays a crucial role alongside orbital hybridization (see Appendix~\ref{app:effective_model}). 
An additional feature of the (001) response is the strong cancellation
between the contributions of different doublets. Indeed, the individual
doublet contributions can become considerably larger than the total
conductivity, showing that the final response results from a delicate
balance between quantum-geometric contributions of opposite sign.
Smaller structures are also visible at lower filling. In particular, the
feature around the onset of the second doublet is naturally associated with the opening of an additional occupied band and the corresponding change in the available phase space. Instead, at lower chemical potentials (see the inset in Fig. ~\ref{fig:band_structure}(c)), where only the first doublet is occupied, a small but clear feature emerges. This behavior is reminiscent of the analogous two-band Rashba system in the presence of an in-plane magnetic field along $\hat{y}$.
In addition, the order of magnitude of the nonlinear conductivity obtained in our calculations is consistent with the wide range of values reported in the literature for different materials~\cite{wang2023quantum, 10.1021/acs.nanolett.5c00195, sala2026, kaplan2024unification, 10.1021/acs.nanolett.5c02968, gao2023quantum, wang2021intrinsic, TIEN2025100027}.

\begin{figure}[t!]
\centering
\includegraphics[width=0.49\textwidth]{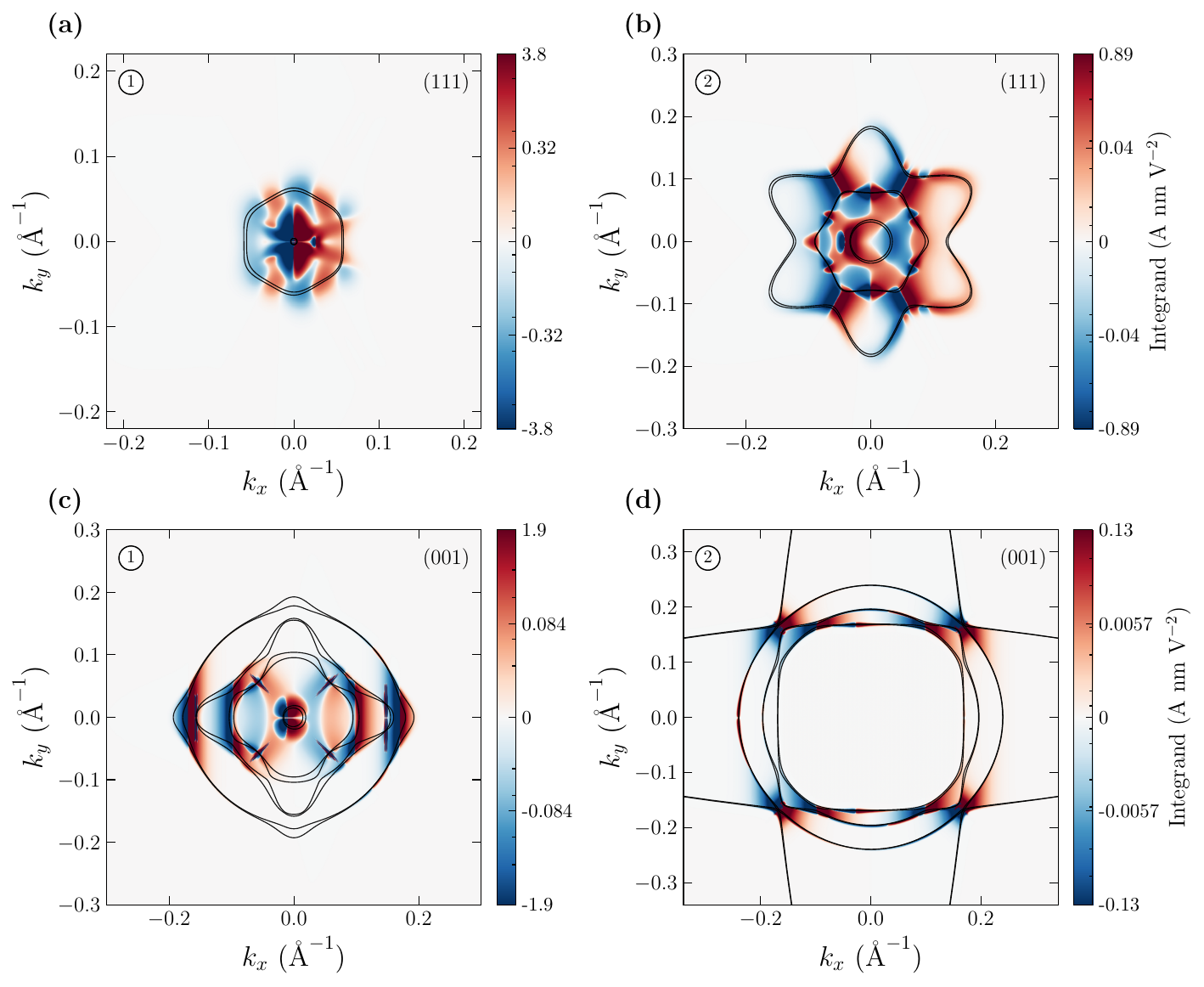}
\caption{
Momentum-resolved integrand, summed over all bands, of Eq.~\eqref{eq:nonlinearQG} for the chemical potentials labeled 1 and 2 in Fig.~\ref{fig:band_structure}. 
Panels (a) and (b) refer to the (111) LAO/STO interface at $\mu=0.0058$ eV and $\mu=0.0214$ eV, respectively, while panels (c) and (d) show the corresponding results for the (001) interface at $\mu=0.1641$ eV and $\mu=0.3171$ eV.
The black contours indicate the Fermi surfaces at the selected chemical potentials. 
The color scale represents the local momentum-space contribution to $\sigma_{xxx}^{\rm QG}$.
All calculations are performed for $B_y=5\,\mathrm{T}$ and $T=10\,\mathrm{K}$.}
\label{fig:integrands}
\end{figure}

The momentum-resolved integrand of Eq.~\eqref{eq:nonlinearQG}, shown in
Fig.~\ref{fig:integrands}, highlights the momentum-space origin of the nonlinear response. For the (111) interface, the momentum-space pattern reflects the threefold
arrangement of the inter-doublet avoided crossings. For the
${xxx}$ component considered here, the strongest hot spots are
associated with the oblique crossing directions, approximately along the
$\pi/3$ and $2\pi/3$ lines in the $k_x$--$k_y$ plane and their symmetry-related counterparts. Their
weight and sign change as different portions of these avoided crossings are
sampled by the Fermi surface, giving rise to the nonmonotonic filling
dependence of the integrated response. For the (001) interface, the strongest hot spots develop around the avoided crossings along
the Brillouin-zone diagonals and contribute significantly when they are
intersected by the Fermi surface. These directions are particularly relevant
because the $d_{yz}$ and $d_{zx}$ dispersions are almost degenerate along
$k_x=k_y$. Their symmetric combination then hybridizes directly with the
$d_{xy}$ state, producing strong interorbital mixing together with a small
avoided gap. Both ingredients enhance the BNQM and lead to the sharp
momentum-space structures visible in Figs.~\ref{fig:integrands}(c) and
\ref{fig:integrands}(d). Away from these regions the integrand rapidly
decreases.
Importantly, at finite magnetic field the dominant matrix elements of the BNQM may
connect the two branches of the same split doublet. However, as shown
analytically in Appendix~\ref{app:effective_model}, these intra-doublet
contributions are themselves generated by the multiorbital hybridization with
the neighboring doublets. The large nonlinear response is therefore
ultimately rooted in the interorbital avoided-crossing structure rather than
in an isolated Rashba splitting.

\begin{figure}[t]
\centering
\includegraphics[width=0.49\textwidth]{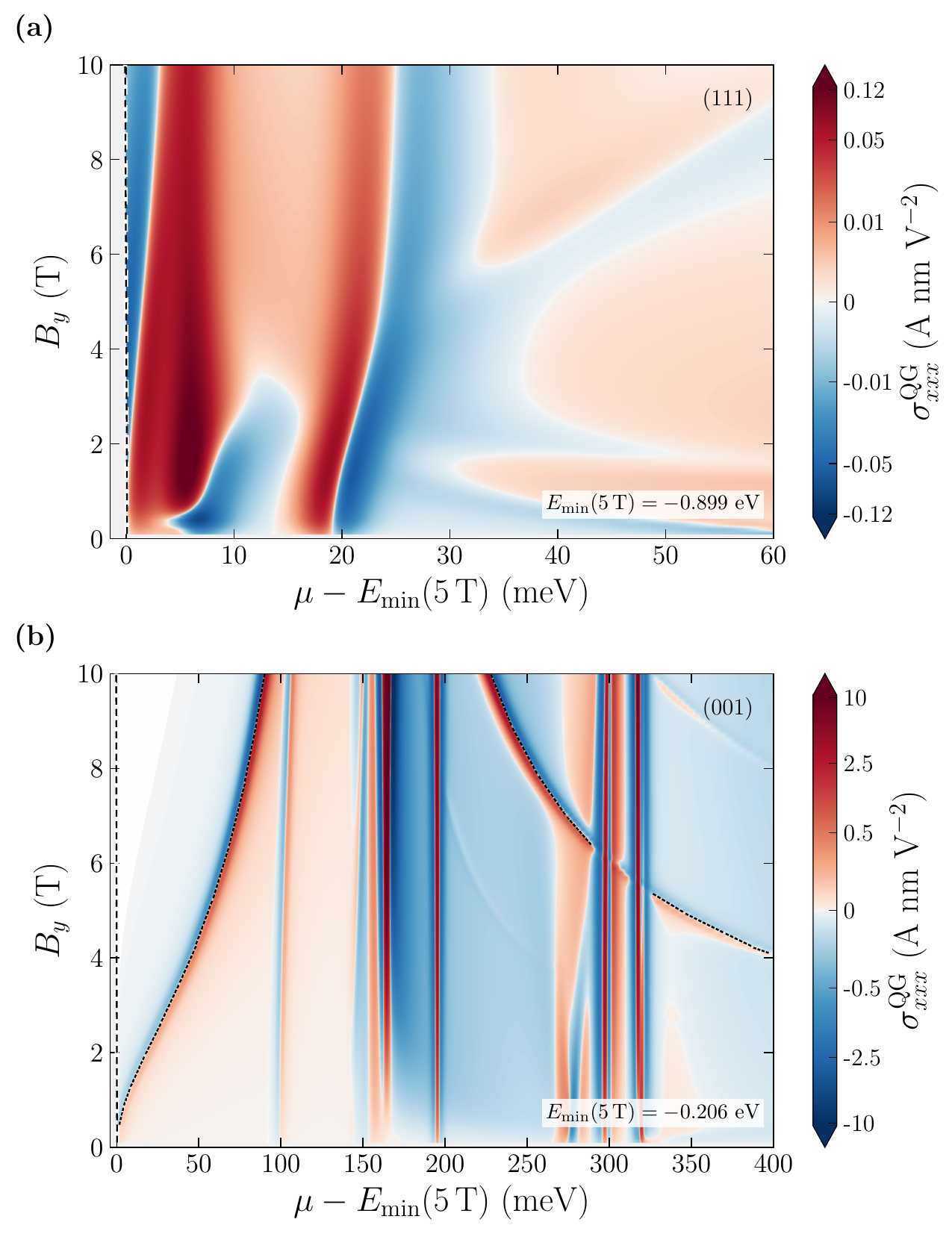}
\caption{
Quantum-geometric contribution to the nonlinear conductivity,
$\sigma_{xxx}^{\rm QG}$, as a function of the in-plane magnetic field
$B_y$ and chemical potential for the (a) (111) and (b) (001) LAO/STO
interfaces. The chemical potential is measured with respect to the
minimum of the lowest band at $B_y=5\,\mathrm{T}$, with $E_{\rm min}(5\,\mathrm{T})=-0.900$ eV for the (111) interface and 
$E_{\rm min}(5\,\mathrm{T})=-0.206$ eV for the (001) interface.
The color scale represents $\sigma_{xxx}^{\rm QG}$ in
A nm V$^{-2}$. 
}  
\label{fig:magnetic_field_behaviour}
\end{figure}

Finally, we investigate the evolution of the quantum-geometric nonlinear
conductivity with the external magnetic field. Figure~\ref{fig:magnetic_field_behaviour}
shows $\sigma^{\rm QG}_{xxx}$ as a function of $B_y$ and chemical potential,
measured with respect to the band minimum at $B_y=5$ T.

The two interfaces display qualitatively different field dependences. 
The (111) interface shows a markedly stronger field dependence (Fig.~\ref{fig:magnetic_field_behaviour}(a)). The boundaries
between positive and negative response shift substantially with $B_y$, and at
fixed chemical potential the conductivity can change sign more than once.
This sensitivity reflects the small SOC, orbital-Rashba, and inter-doublet
energy scales controlling the avoided crossings, which become comparable to
the Zeeman energy already at moderate fields. As a result, the magnetic field
strongly reshapes the local hybridization pattern and the associated
quantum-geometric response.
For the
(001) interface, the narrow peak-and-dip structures appear as nearly vertical
stripes over a broad field range (Fig.~\ref{fig:magnetic_field_behaviour}(b)). Their position is therefore mainly
controlled by the chemical potential and remains only weakly affected by the
Zeeman field. In the (001) case, this behavior reflects the fact that, in the magnetic-field range considered here, the Zeeman splitting remains smaller than the Rashba one. As a result, the overall profile of the nonlinear conductivity remains largely unchanged as the magnetic field varies, leading to the nearly vertical stripe-like structures observed. Two more curved branches (see dashed lines in Fig.~\ref{fig:magnetic_field_behaviour}(b)), however, can be identified as exceptions; one of them is found at low filling and can be traced back to the contribution of the first doublet.

We also notice that the evolution of the sign-changing regions does not simply follow
the shift of the band minimum, showing that the modulation cannot be
understood as a rigid displacement of the band structure.

Therefore, the (111) response provides a field-tunable nonlinear signal,
whereas the robust peak-and-dip structure of the (001) interface offers an
experimental marker of the Lifshitz transitions as the carrier density is
varied.

\section{Summary}
In conclusion, we have theoretically investigated the role of quantum geometry in determining the intrinsic non-linear conductivity of STO-based 2DEGs. By comparing the prototypical (001) and (111) interface orientations within a multi-orbital tight-binding model, we have discussed how accounting for the multiorbital nature of Bloch electrons is essential to accurately evaluate the quantum geometric tensor and predict the non-linear transport properties. 
In particular, we show how the interplay between the multiband electronic structure and orbital hybridization near avoided band crossings shapes the momentum-space distribution of the band-normalized quantum metric and the resulting non-linear transport response.
A conclusion of our study is the strong dependence of the non-linear response on the crystal orientation. For the (001) interface, the geometric contribution to the non-linear conductance is dominated by sharp peak-and-dip features centered around Lifshitz transitions, while remaining negligible in the low-doping regime. In contrast, the trigonal symmetry of the (111) interface enhances the mixing between $a_{1g}$ and $e_{\pi}$ orbitals. This leads to a non-monotonic behavior of the non-linear response, maintaining a significant quantum geometric contribution over a much broader range of carrier densities. 
Our work establishes the tuning of crystallographic orientation and the avoided-crossing engineering as powerful tools towards controlling intrinsic non-linear transport phenomena in oxide heterostructures, opening new pathways for quantum-geometry-based electronics applications. 

\section{Acknowledgments}
I.G., M.T. and R.C. acknowledge F. Romeo, C. Guarcello, A. Maiellaro, and R. Capecelatro for fruitful discussions.
M.T. and R.C. acknowledge funding from IQARO (Spinorbitronic Quantum Bits in Reconfigurable 2D Oxides) project of the European Union’s Horizon Europe research and innovation programme under grant agreement n. 101115190.
This work was funded by the Deutsche Forschungsgemeinschaft (DFG, German Research Foundation) – 545818886; 328545488—TRR 227, Project B13. I.G. acknowledges financial support from the Erasmus+ Traineeship programme during her doctoral research stay under the supervision of A.J.

\newpage
\onecolumngrid

\appendix
\section{Hamiltonian details}\label{app:hamiltonians}
\subsection{(001) Interface}
The theoretical framework adopted for the (001)-STO based 2DEG follows the formulation presented in Refs.~\cite{khalsa2013theory, zhong2013theory, vivek2017topological, vaz2019mapping}, and is briefly reviewed in the following. The electronic structure is obtained from a tight-binding Hamiltonian defined on a square lattice of Ti atoms and includes the two lowest-energy $d_{xy}$ bands together with the $d_{yz}$ and $d_{zx}$ bands. In the basis $(d_{xy1},d_{xy2},d_{yz},d_{zx})\otimes(\uparrow,\downarrow)$, the Hamiltonian takes the form:
\begin{equation}
    H= H_{TB}(t,t_h) + H_{SOC}(\lambda)+ H_{OM}(f)
\end{equation}
$H_{TB}= \hat{h}_{TB} \otimes \sigma_0$ is the diagonal term for the nearest-neighbour hopping (quantified by $t,t_h$) with the diagonal elements of $\hat{h}_{TB}$
\begin{equation}
\begin{aligned}
\epsilon_{xy_j} &= 2t_l\left[2-\cos(k_xa)-\cos(k_ya)\right]+o_j,\\
\epsilon_{yz} &= 2t_l\left[1-\cos(k_ya)\right]
                +2t_h\left[1-\cos(k_xa)\right]+o_3,\\
\epsilon_{zx} &= 2t_l\left[1-\cos(k_xa)\right]
                +2t_h\left[1-\cos(k_ya)\right]+o_3.
\end{aligned}
\label{eq:diagonal_elements}
\end{equation}
$H_{SOC}= (2\lambda \bf L\cdot S)$ represents the spin-orbit interaction $\lambda$, where $\mathbf L=(L_x,L_y,L_z)$ denotes the orbital angular momentum matrices expressed in the orbital basis and $\mathbf S=\boldsymbol{\sigma}/2$ is the electron spin operator, with $\boldsymbol{\sigma}$ representing the Pauli matrices (in units of $\hbar=1$). $H_{OM}= \hat{h}_{OM}  \otimes \sigma_0 $ includes the orbital mixing $f_{jm}=2ig_j\sin{(k_m a)}$ where $\hat{h}_{OM}= \{(0,0,f_{1x},f_{1y}),(0,0,f_{2x}, f_{2y}), (-f_{1x},-f_{2x}, 0,0),(-f_{1y},-f_{2y}, 0,0)\}$.\\
To investigate the non linear conductivity driven by the quantum geometry, it is necessary to introduce a planar magnetic field in our system. So, we consider two additional terms $H^{spin}_Z + H^{orb}_Z$ represent the coupling of the external magnetic field with spin and orbital degrees of freedom. In particular, the spin and orbital Zeeman contributions are described by the terms
$H_Z^{\mathrm{spin}}=\frac{g_s\mu_B B_y}{2}\left(\mathbb{I}_{4}\otimes\sigma_y\right)$
and
$H_Z^{\mathrm{orb}}=g_L\mu_B B_y (L_y\otimes \sigma_0)$,
where $g_s=2$ and $g_L=1$.

The model parameters are $t_l = 0.388~\mathrm{eV}$, $t_h = 0.031~\mathrm{eV}$, $o_1 = -0.205~\mathrm{eV}$, $o_2 = -0.105~\mathrm{eV}$, $o_3 = -0.0544~\mathrm{eV}$, $g_1 = 0.002~\mathrm{eV}$, $g_2 = 0.005~\mathrm{eV}$, and $\lambda = 1/120~\mathrm{eV}$, yielding good agreement with the ARPES measurements reported in Ref.~\cite{vaz2019mapping}. Here, $o_j$ are the orbital onsite energies and $a = 3.9\,\text{\AA}$ is the lattice constant.

\subsection{(111) Interface}
We describe the (111) STO-based 2DEG using a minimal tight-binding model
for two adjacent Ti layers, following
Refs.~\cite{trama2021straininduced,trama2023effect,trama2026inhomogeneous}.
Retaining the three $t_{2g}$ orbitals $(d_{yz},d_{zx},d_{xy})$ and spin on
each layer gives a twelve-band Hamiltonian,
\begin{equation}
H=H_{\mathrm{TB}}+H_{\mathrm{SOC}}+H_{\mathrm{TRI}}+H_{\nu},
\qquad H_{\nu}=H_v+H_R.
\end{equation}
The orbital-conserving hopping term $H_{\mathrm{TB}}$ contains the direct
and indirect interlayer amplitudes $t_D=0.50~\mathrm{eV}$ and
$t_I=0.04~\mathrm{eV}$, respectively. These are the bilayer parameters
obtained in Ref.~\cite{trama2021straininduced} by fitting the ARPES data
of Ref.~\cite{mckeown2014control}; orbital-conserving hopping within
each layer is neglected.

The atomic spin--orbit interaction is
$
H_{\mathrm{SOC}}=-\lambda\,\mathbf L\cdot\mathbf S
$ with $\lambda=0.01~\mathrm{eV}$,
where $\mathbf S=\boldsymbol\sigma/2$ and $\boldsymbol\sigma$ denotes
the spin Pauli matrices.\footnote{For the (111) model, $\mathbf L$
denotes the dimensionless $l=1$ matrices in the
$(d_{yz},d_{zx},d_{xy})$ basis, with
$(L_\alpha)_{\beta\gamma}=-i\epsilon_{\alpha\beta\gamma}$ and
$(\alpha,\beta,\gamma)\in\{x,y,z\}$.
The physical atomic orbital angular momentum projected onto the
$t_{2g}$ manifold is $-\hbar\mathbf L$.}
Identity operators in the layer and spin spaces are left implicit
where appropriate. The trigonal crystal field is
\begin{equation}
H_{\mathrm{TRI}}=\frac{\Delta}{2}
\begin{pmatrix}0&1&1\\1&0&1\\1&1&0\end{pmatrix}
=\Delta\mathbb I_3-\frac{3\Delta}{2}L_{111}^{,2},
\qquad \Delta=-0.005~\mathrm{eV},
\end{equation}
with $L_{111}=(L_x+L_y+L_z)/\sqrt3$.
Before SOC is included, this term gives energies $\Delta$ and
$-\Delta/2$ to the $a_g=(d_{yz}+d_{zx}+d_{xy})/\sqrt3$ singlet and
the $e_g^\pi$ doublet, respectively. Thus $\Delta<0$ lowers the
$a_g$ orbital, as appropriate for the trigonal compression considered
here.

The interfacial electrostatic potential makes the two layers inequivalent
and breaks inversion symmetry. Its onsite contribution is
\begin{equation}
H_v=\operatorname{diag}(v_1,v_2)\otimes\mathbb I_6,
\qquad (v_1,v_2)=(0.10,0.20)~\mathrm{eV}.
\end{equation}
The same electric field induces spin-independent interorbital hopping,
which gives the orbital Rashba contribution. We retain its full lattice
form, including both intra- and interlayer processes. For the potential
profile used here, its orbital blocks are
\begin{equation}
H_R(\mathbf k)=
\begin{pmatrix}
r_{\parallel}(\mathbf k)&r_{\perp}(\mathbf k)\\
r_{\perp}^{\dagger}(\mathbf k)&r_{\parallel}(\mathbf k)
\end{pmatrix},
\end{equation}
with the spin identity understood and $\mathbf k$ expressed in units of $1/\tilde{a}$
where $\tilde{a}=\sqrt{2/3}\,a_0$ and $a_0=3.905~\text{\AA}$.
To specify these blocks compactly, define
\begin{equation}
\mathcal A(u,v,w)=\begin{pmatrix}0&u&v\\-u&0&w\\-v&-w&0\end{pmatrix}.
\end{equation}
Then
\begin{align}
r_{\parallel}&=\frac{2i\eta_p}{\sqrt{15}}
\left[V_{pd\pi}\mathcal A(\mathbf f)
+\sqrt3 V_{pd\sigma}\mathcal A(\mathbf g)\right],\\
r_{\perp}&=\frac{2^{7/4}\eta_p V_{pd\pi}}{\sqrt{15}}
\mathcal A(z_2-z_1,z_2-z_3,z_1-z_3),
\end{align}
where $\mathcal A(\mathbf f)$ denotes $\mathcal A(f_1,f_2,f_3)$, and
\begin{align}
\mathbf f&=(-2s_0+s_-+s_+,\;-2s_++s_0-s_-,\;2s_--s_0+s_+),\\
\mathbf g&=(-s_--s_+,\;-s_0+s_-,\;s_0-s_+),\\
s_0&=\sin(\sqrt3q_x),\qquad
s_{\pm}=\sin\!\left(\frac{\sqrt3q_x\pm3q_y}{2}\right),\\
z_1&=e^{i(\sqrt3q_x-q_y)/2},\qquad
z_2=e^{-i(\sqrt3q_x+q_y)/2},\qquad z_3=e^{iq_y}.
\end{align}
We use $V_{pd\pi}=0.028~\mathrm{eV}$,
$V_{pd\sigma}=-0.064~\mathrm{eV}$ and the dimensionless field-induced
orbital-mixing parameter $\eta_p\simeq0.0443547$.
The sine terms are odd in momentum, while the phase factors in the
interlayer block retain the full bilayer geometry. Together with atomic
SOC, these terms generate the Rashba spin splitting.

Finally, the magnetic field used in the main text is included through
\begin{equation}
H_Z=\mu_B\mathbf B\cdot(-\mathbf L+2\mathbf S),
\end{equation}
consistently with the orbital convention stated above.

\section{Electronic density}\label{app:density}
\begin{figure}[H]
\centering
\includegraphics[width=0.9\textwidth]{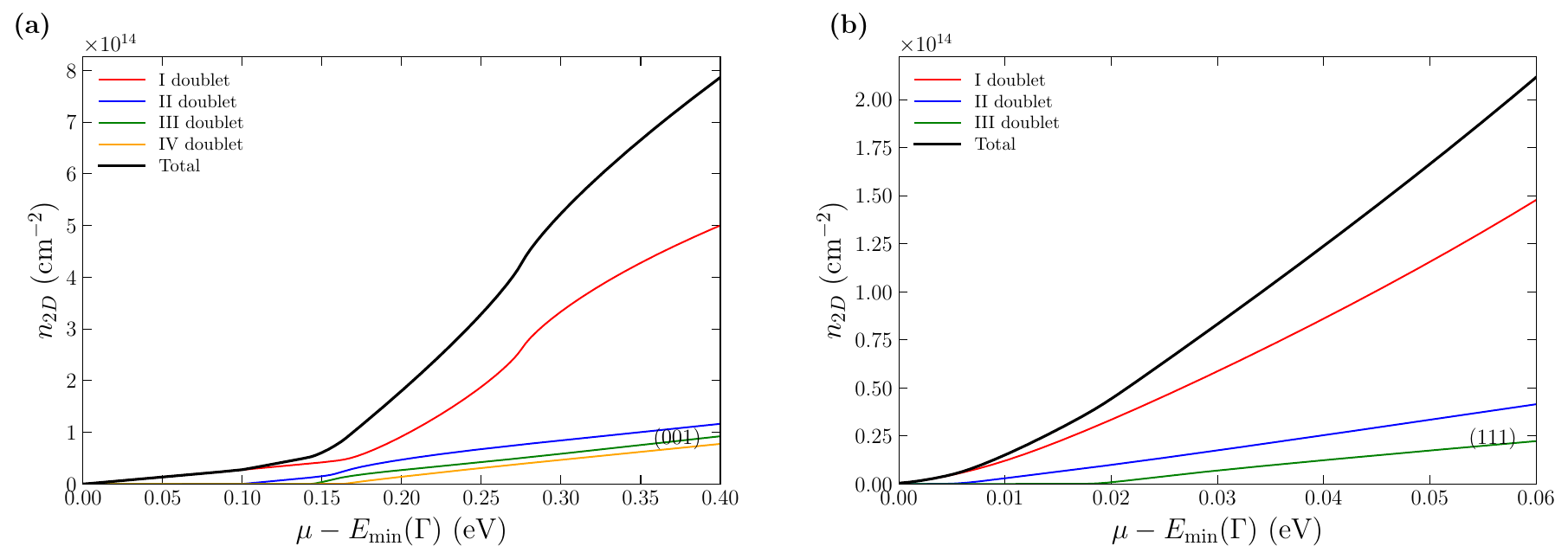}
\caption{
Two-dimensional electron density, $n_{2D}$, as a function of the chemical potential for the (a) (001) and (b) (111) LAO/STO interfaces. The black curves represent the total electron density, while the colored curves indicate the contributions from the individual band doublets. The chemical potential is measured from the minimum of the lowest band at $\Gamma$, $E_{\rm min}(\Gamma)$. The electron density is obtained by integrating the Fermi--Dirac occupation over the Brillouin zone. Calculations are performed at $B_y=5\,\mathrm{T}$ and $T=10\,\mathrm{K}$.
}  
\label{fig:electronic density}
\end{figure}

To complement the analysis of the quantum-geometric nonlinear conductivity presented in the main text, we investigate the two-dimensional electron density as a function of the chemical potential for both LAO/STO interface orientations.
Figure~\ref{fig:electronic density} shows the total electron density together with the contributions from the individual band doublets. For both orientations, the total density increases monotonically
with the chemical potential, reflecting the progressive filling of the electronic bands.\\
For the (001) interface (Fig.~\ref{fig:electronic density}(a)), the second, third, and fourth doublets start contributing at approximately 0.10, 0.14, and 0.16 eV above the lowest band minimum, respectively. These thresholds are associated with the appearance of additional Fermi-surface pockets, characteristic of band-filling Lifshitz transitions. The corresponding changes in the slope of the electron density reflect variations in the density of states. However, the pronounced peak-and-dip structures of the quantum-geometric conductivity in Fig.~\ref{fig:band_structure}(c) do not simply follow these occupation thresholds. As discussed in the main text, their enhancement is primarily associated with interorbital hybridization
near the avoided crossings of the band structure
(Fig.~\ref{fig:band_structure}(d)).\\
For the (111) interface (Fig.~\ref{fig:electronic density}(b)), the total electron
density is predominantly determined by the lowest doublet
throughout the investigated filling range.
The second and third doublets become occupied at
approximately 0.006 and 0.019 eV, respectively,
signaling the emergence of additional Fermi-surface pockets. These energy scales are particularly relevant when
compared with the nonlinear conductivity in Fig.~\ref{fig:band_structure}(a).
At low fillings, the first doublet provides the dominant contribution to the quantum-geometric response, while the second and third doublets become increasingly
important as the chemical potential increases.
In particular, the higher-filling peak, dominated by the third doublet, is also captured by the effective two-band description discussed in the main text.

Interestingly, while the electron density exhibits a
rather regular increase with chemical potential, the quantum-geometric conductivity displays a strongly nonmonotonic behavior.
This comparison highlights that the nonlinear response is not simply determined by the occupation of the individual doublets, but is strongly influenced by
the interband matrix elements and the multiorbital
hybridization near avoided crossings.

\section{Effective model near degeneracies}\label{app:effective_model}
To understand the quantum-geometric structure close to the degeneracy points
of the (111) interface, we consider an expansion around the first avoided
crossing between the first and second doublets at $k_y\simeq0$ and study
$\Lambda_{yyy}(k_x,k_y\simeq0)$ at $B=0$. We focus on $\Lambda_{yyy}$ because it probes the momentum dependence
transverse to the avoided crossing at $k_y=0$, allowing for a simple analytical treatment of the local quantum-geometric structure.
In this case the integrated
BNQM-dipole contribution vanishes because of time-reversal symmetry, while
the local momentum-space structure can still be analyzed. The crossing
considered here is representative of the symmetry-related avoided crossings
which generate the quantum-geometric hot spots discussed in the main text.

Considering first the tight-binding Hamiltonian, the crossing occurs between
the bonding orbitals
$\sqrt{2}\ket{d_{yz}^-}=\ket{d_{yz}^1}-\ket{d_{yz}^2}$ and
$\sqrt{2}\ket{d_{zx}^-}=\ket{d_{zx}^1}-\ket{d_{zx}^2}$.
We therefore restrict the analysis to the subspace
$\{\ket{d_{yz}^-\uparrow},\ket{d_{zx}^-\uparrow},
\ket{d_{yz}^-\downarrow},\ket{d_{zx}^-\downarrow}\}$.
For simplicity, in the following we drop the label "$-$" denoting the bonding
orbitals and neglect the layer-dependent on-site contribution.
Expanding around $k_y\simeq0$, the Hamiltonian in this subspace reads
\begin{equation}
    H_{\rm eff}(k_x,k_y\simeq0)
    =
    \phi(k_x)k_y\tau_z\sigma_0
    +\xi(k_x)\tau_y\sigma_0
    -\frac{\lambda}{2}\tau_y\sigma_z
    +\frac{\Delta}{2}\tau_x\sigma_0 ,
    \label{eq:expansion}
\end{equation}
where $\tau_i$ are Pauli matrices acting in the
$\{\ket{d_{yz}},\ket{d_{zx}}\}$ orbital space. The first term is obtained by
expanding the tight-binding Hamiltonian to first order in $k_y$, while the
orbital Rashba term is evaluated at $k_y=0$. The last two terms describe the
atomic SOC and the trigonal crystal field, respectively. Terms proportional
to $\tau_0\sigma_0$ have been subtracted.

The expanded coefficients are
\begin{equation}
    \phi(k_x)
    =
    \frac{t_D-t_I}{2}
    \sin\left(\frac{\sqrt{3}}{2}k_x\right)
\end{equation}
and
\begin{equation}
    \xi(k_x)
    =
    \frac{8\eta_p}{\sqrt{15}}
    \sin\left(\frac{\sqrt{3}}{4}k_x\right)
    \left[
        -2p_1\cos\left(\frac{\sqrt{3}}{4}k_x\right)
        +(p_1+p_3)
        \cos\left(\frac{3\sqrt{3}}{4}k_x\right)
    \right].
\end{equation}

Hamiltonian~\eqref{eq:expansion} can be diagonalized analytically since the
two spin sectors are decoupled. Defining
\begin{equation}
    M_\sigma^2
    =
    \Delta^2+(2\xi-\sigma\lambda)^2 ,
\end{equation}
the eigenvalues are
\begin{equation}
    \varepsilon_{s,\sigma}
    =
    sE_\sigma,
    \qquad
    E_\sigma
    =
    \frac{1}{2}
    \sqrt{4\phi^2k_y^2+M_\sigma^2},
    \qquad
    s=\pm1 .
\end{equation}

To compute $G_{yy}$, we evaluate the matrix elements of
$\hat v_y=\partial_{k_y}H_{\rm eff}$ between the two states belonging to the
same spin sector. Their modulus is
\begin{equation}
    \left|
    \langle s,\sigma|\hat v_y|-s,\sigma\rangle
    \right|
    =
    \frac{\phi M_\sigma}{2E_\sigma}.
\end{equation}
Since $\hat v_y$ does not mix the two spin sectors, the corresponding BNQM is
\begin{equation}
    G_{yy}^{s,\sigma}
    =
    s\frac{\phi^2M_\sigma^2}{16E_\sigma^5}.
    \label{eq:Gyy_local}
\end{equation}
Taking the derivative with respect to $k_y$ gives
\begin{equation}
    \Lambda_{yyy}^{s,\sigma}
    =
    \partial_{k_y}G_{yy}^{s,\sigma}
    =
    -s\frac{5\phi^4M_\sigma^2}{16E_\sigma^7}k_y .
    \label{eq:Lyyy_local}
\end{equation}

Equation~\eqref{eq:Lyyy_local} directly shows the enhancement of the
quantum-geometric contribution close to a small avoided gap. Introducing the
momentum scale
\begin{equation}
    k_c^\sigma=\frac{M_\sigma}{2|\phi|},
\end{equation}
two regimes can be distinguished. Very close to the center of the avoided
crossing, $|k_y|\ll k_c^\sigma$, the finite gap dominates and
\begin{equation}
    \Lambda_{yyy}^{s,\sigma}\propto k_y .
\end{equation}
For a sufficiently small gap, an intermediate regime
$k_c^\sigma\ll|k_y|$ can instead be reached within the validity of the local
expansion. In this regime $E_\sigma\simeq|\phi k_y|$ and
\begin{equation}
    \Lambda_{yyy}^{s,\sigma}
    \propto
    \operatorname{sgn}(k_y)|k_y|^{-6}.
\end{equation}
This strong enhancement gives rise to the
sharp quantum-geometric hot spots observed in momentum space.

\subsection*{Finite magnetic field}

We now consider a finite magnetic field along $x$ and the corresponding $\Lambda_{yyy}$ response, symmetry-related to the $B_y$, $\Lambda_{xxx}$ configuration considered in the main text. 
Since the orbital Rashba and Zeeman energies have the same order of magnitude, we treat them at the same
perturbative order and write $H_{\rm eff}=H_0+V$, with
\begin{equation}
    H_0=
    \phi k_y\tau_z\sigma_0
    -\frac{\lambda}{2}\tau_y\sigma_z
    +\frac{\Delta}{2}\tau_x\sigma_0,
    \qquad
    V=
    \xi\tau_y\sigma_0
    +b_x\tau_0\sigma_x ,
\end{equation}
where $b_x=g_s\mu_BB_x/2$. The unperturbed spectrum is
$E_s=sE$, with
$E=\sqrt{4\phi^2k_y^2+\lambda^2+\Delta^2}/2$ and $s=\pm1$.
Each level is twofold degenerate.

We first diagonalize $V$ within each degenerate subspace. Introducing Pauli
matrices $\rho_i$ acting within this two-dimensional subspace, the projected
perturbation can be written as
\begin{equation}
    P_sVP_s
    =
    \boldsymbol{h}_s\cdot\boldsymbol{\rho}
    =
    d\,\boldsymbol{n}_s\cdot\boldsymbol{\rho},
    \qquad
    \boldsymbol{n}_s=\frac{\boldsymbol{h}_s}{d},
\end{equation}
where $\boldsymbol{n}_s$ is a unit vector and
\begin{equation}
    d=|\boldsymbol{h}_s|
    =
    \frac{
        \sqrt{
            \lambda^2\xi^2+
            b_x^2(\Delta^2+4\phi^2k_y^2)
        }
    }{2E}.
\end{equation}
The eigenstates of the projected perturbation are denoted by
$\ket{s,\eta}$, with
\begin{equation}
    \boldsymbol{n}_s\cdot\boldsymbol{\rho}
    \ket{s,\eta}
    =
    \eta\ket{s,\eta},
    \qquad
    \eta=\pm1 ,
\end{equation}
and their energies are
$\varepsilon_{s,\eta}=sE+\eta d$.

The velocity operator is $\hat v_y=\phi\tau_z$. Defining
\begin{equation}
    c=\boldsymbol{n}_+\cdot\boldsymbol{n}_- ,
\end{equation}
the overlap between the pseudospin eigenstates of the two manifolds is
\begin{equation}
    \left|
    \langle\boldsymbol{n}_+,\eta|
    \boldsymbol{n}_-,\eta'\rangle
    \right|^2
    =
    \frac{1+\eta\eta'c}{2}.
\end{equation}
Using
$v_0^2=\phi^2(\lambda^2+\Delta^2)/(4E^2)$, the inter-doublet contribution
to the BNQM reads
\begin{equation}
    G_{yy,{\rm inter}}^{s,\eta}
    =
    v_0^2
    \sum_{\eta'=\pm1}
    \frac{1+\eta\eta'c}
    {\left(2sE+\eta d-\eta'd\right)^3}.
    \label{eq:Gyy_inter_finiteB}
\end{equation}
For $d\ll E$,
\begin{equation}
    G_{yy,{\rm inter}}^{s,\eta}
    \simeq
    s\frac{\phi^2(\lambda^2+\Delta^2)}
            {16E^5}
    \left[
        1-s\eta\frac{3d}{2E}(1-c)
    \right].
    \label{eq:Gyy_inter_expanded}
\end{equation}
The $E^{-5}$ dependence shows that the inter-doublet contribution remains
enhanced close to the avoided crossing.

At finite field, an additional contribution arises from transitions between
the two branches $\eta=\pm1$ of the same doublet. At zeroth order such
transitions vanish because
\begin{equation}
    P_s\hat v_yP_s
    =
    s\frac{\phi^2k_y}{E}P_s
\end{equation}
is proportional to the identity within the degenerate subspace. A finite
intra-doublet matrix element appears only through mixing with the opposite
manifold. To first order in $V$,
\begin{equation}
    |\Psi_{s,\eta}\rangle
    =
    |s,\eta\rangle
    +
    \frac{1}{2sE}P_{-s}V|s,\eta\rangle
    +O(V^2),
\end{equation}
and therefore
\begin{equation}
    \langle\Psi_{s,+}|\hat v_y|\Psi_{s,-}\rangle
    =
    \frac{
    \langle s,+|
    \left(
        VP_{-s}\hat v_y+\hat v_yP_{-s}V
    \right)
    |s,-\rangle
    }{2sE}
    +O(V^2).
    \label{eq:vintra}
\end{equation}
The corresponding BNQM contribution is
\begin{equation}
    G_{yy,{\rm intra}}^{s,\eta}
    =
    2
    \frac{
        \left|
        \langle\Psi_{s,\eta}|
        \hat v_y
        |\Psi_{s,-\eta}\rangle
        \right|^2
    }
    {(2\eta d)^3}.
    \label{eq:Gyy_intra}
\end{equation}
Although this transition occurs within the same doublet, its matrix element
is generated by virtual mixing with the opposite doublet. The finite-field
quantum geometry therefore contains a direct inter-doublet contribution
associated with the avoided crossing and an intra-doublet contribution
induced by the same inter-doublet hybridization and enhanced by the small
splitting $2d$.
\subsection*{Generalization to the (001) interface}

A similar mechanism can be identified for the (001) interface. For simplicity,
we consider a single-layer three-orbital model containing the
$\{d_{xy},d_{yz},d_{zx}\}$ orbitals of the first Ti layer, thus excluding the
second $d_{xy}$ band of the full model.

The avoided crossings relevant for the quantum-geometric response occur along
the diagonals of the Brillouin zone. Along $k_x=k_y$, the $d_{yz}$ and
$d_{zx}$ tight-binding dispersions are degenerate. Introducing
$\ket{d_{\pm}}=(\ket{d_{yz}}\pm\ket{d_{zx}})/\sqrt{2}$, the
inversion-breaking orbital mixing couples $d_{xy}$ only to $d_+$, with
$m_R(k)=2\sqrt{2}g\sin k$, while $d_-$ is decoupled from this term. The atomic
SOC provides an additional contribution to the $d_{xy}$--$d_+$ hybridization.

We restrict the analysis to the $\{d_{xy},d_+\}$ sector and introduce
$k_\parallel=(k_x+k_y)/\sqrt{2}$ and
$q=k_\parallel-k_\parallel^\star$, where $k_\parallel^\star$ is the position
of the corresponding bare crossing. After subtracting the term proportional
to the identity and expanding around $q=0$, the effective Hamiltonian reads
\begin{equation}
    H_{{\rm eff},\mu}^{(001)}
    =
    u q\,\tau_z-m_\mu\tau_y ,
    \qquad
    m_\mu=
    2\sqrt{2}g
    \sin\left(\frac{k_\parallel^\star}{\sqrt{2}}\right)
    +\mu\lambda ,
    \label{eq:001_effective}
\end{equation}
where $\mu=\pm1$ labels the two spin-orbit sectors and
$u=(t_l-t_h)\sin(k_\parallel^\star/\sqrt{2})/\sqrt{2}$ is the relative
velocity of the two bare bands. The relative on-site energies determine
$k_\parallel^\star$, while the orbital mixing and SOC determine the mass
$m_\mu$ and the corresponding avoided gap $2|m_\mu|$.

The spectrum is
\begin{equation}
    E_{s,\mu}=sE_\mu,
    \qquad
    E_\mu=\sqrt{u^2q^2+m_\mu^2},
    \qquad s=\pm1 .
\end{equation}
Neglecting the weak momentum dependence of $m_\mu$ close to the crossing,
$\hat v_\parallel=u\tau_z$, and therefore
\begin{equation}
        G_{\parallel\parallel}^{s,\mu}
    =
    s\frac{u^2m_\mu^2}{4E_\mu^5},\; {\implies} \;
    \Lambda_{\parallel\parallel\parallel}^{s,\mu}
    =
    -s\frac{5u^4m_\mu^2}{4E_\mu^7}q .
\end{equation}
As for the (111) interface, the finite gap gives
$\Lambda_{\parallel\parallel\parallel}\propto q$ close to the center of the
avoided crossing, $|q|\ll |m_\mu/u|$, while for a sufficiently small gap the
intermediate regime $|m_\mu/u|\ll |q|$ gives
$\Lambda_{\parallel\parallel\parallel}\propto
\operatorname{sgn}(q)|q|^{-6}$.

The quantum-geometric hot spots of the (001) interface can therefore be traced
back to the interorbital avoided crossings between the $d_{xy}$ and
$d_{yz}/d_{zx}$ sectors close to the Lifshitz transition. Along the opposite
diagonal, $k_x=-k_y$, the same argument applies with $d_+$ replaced by $d_-$.

\twocolumngrid
\bibliography{Bib}

\end{document}